\documentclass[runningheads]{llncs}
\usepackage[T1]{fontenc}

\usepackage{cite}
\usepackage{amsmath,amssymb,amsfonts}
\usepackage{algorithmic}
\usepackage{graphicx}
\usepackage{textcomp}
\usepackage{xcolor}
\usepackage{multirow}

\usepackage{orcidlink}
\usepackage{lipsum}
\usepackage[ruled,vlined]{algorithm2e}

\begin{document}


\title{On the Costs and Benefits\\of Dynamizing a Learned Index}

\author{Terézia Slanináková\inst{1,2}\orcidID{0000-0003-0502-1145} \and Jaroslav Olha\inst{1}\orcidID{0000-0003-1824-468X} \and David Proch\'azka\inst{1}\orcidID{0009-0000-2765-8329} \and Matej Antol\inst{1,2}\orcidID{0000-0002-1380-5647} \and Vlastislav Dohnal\inst{1}\orcidID{0000-0001-7768-7435}}

\authorrunning{T. Slanináková et al.}

\institute{Faculty of Informatics, Masaryk University, Brno, Czechia \and Institute of Computer Science, Masaryk University, Brno, Czechia}





\maketitle

\begin{abstract} 
One of the main challenges within the growing research area of learned indexing is the lack of adaptability to dynamically expanding datasets. This paper explores the dynamization of a static learned index for complex data through operations such as node splitting and broadening, enabling efficient adaptation to new data. Furthermore, we evaluate the trade-offs between static and dynamic approaches by introducing 
an amortized cost model to assess query performance in tandem with the build costs of the index structure, enabling experimental determination of when a dynamic learned index outperforms its static counterpart. We apply the dynamization method to a static learned index and demonstrate that its superior scaling quickly surpasses the static implementation in terms of overall costs as the database grows.

\end{abstract}

\section{Introduction}
The problem of adapting to dynamically expanding datasets remains a challenge in many indexing approaches. For instance, many recent advances in indexing involve machine learning models, leading to the emergence of a new specialized field of study called \textit{learned indexing}. Once trained, machine learning models typically cannot be updated with new data or classification categories without losing prior knowledge, often requiring a full retraining instead.

Despite this limitation, learned indexing has proven successful at indexing structured, low-dimensional datasets~\cite{kraska2018case,ding2020alex,wu2021updatable}, and continues to gain traction in complex data indexing and retrieval~\cite{dong2020learning,antol2021learned,gupta2022bliss,lampropoulos2023adaptive,li2024flex}. However, when it comes to indexing complex, high-dimensional data, the models also tend to become more complex~\cite{li2024flex}.

It is reasonable to assume that learned models are on their way to modeling complex data distributions more efficiently than traditional indexing methods, raising the question of the future role of indexing experts in such a transition. One option is for indexing experts to develop machine learning expertise and design models that better support dynamic learning and effectively manage complex data. Another, more feasible option -- though not mutually exclusive -- is to address the indexing questions separately from the core machine learning challenges, making improvements to learned indexing methods at a higher level of abstraction. This approach would be analogous to how indexing research rarely concerns itself with the physical characteristics of storage media, leaving those lower levels of abstraction to storage systems engineers.

This paper is focused on the latter approach, proposing a dynamization method capable of transforming a static learned indexing structure into a dynamic one through generalized node-splitting and node-broadening operations, as well as a set of simple rules for their usage. Even though it was developed with a learned indexing use case in mind, this approach is general enough to be applicable to any partition-based indexing technique that struggles with dynamization.

A key challenge, however, is the evaluation of the costs and benefits of such a dynamization, since it can be difficult to compare an index with high upfront construction costs to one where those costs are distributed more evenly over the lifetime of the database.

Thus, we compare the static and dynamic approaches in terms of their amortized costs. We do this by establishing several indexing scenarios, and defining them in a way that allows us to quantify the amortized build costs of an average query, in addition to its search costs. This results in a single amortized cost metric that can be used to evaluate, in a very clear and straightforward manner, when dynamization is appropriate and when it might be counter-productive.

\section{Related Work}
\label{sec:relatedwork}

This paper focuses on dynamic learned indexing of complex data -- while research specifically on this topic is limited, both learned indexing and dynamic indexing have been extensively studied as separate topics. Additionally, the idea of dynamizing static indexes in an index-agnostic manner has been explored in prior research, which we also review in this section.

\subsection{Learned Indexing}

Learned indexing~\cite{kraska2018case,learnedIndexesSurvey} has emerged as one of the novel approaches to indexing, based on the premise that indexes can be viewed as models~\cite{kraska2018case}. However, the straightforward idea of learning cumulative distribution function for one-dimensional data is not directly applicable to high-dimensional data. Most of the existing learned indexes for high-dimensional data either create their own clustering~\cite{dong2020learning,slaninakova2021data,gupta2022bliss,li2023learning} or learn an existing one~\cite{antol2021learned,li2024flex}. The limitation of such approaches is that they are designed to index static datasets only. They do not address the dynamic indexing scenario, aside from resorting to a complete rebuild of the index~\cite{li2024flex} without any guiding policy, essentially relying on the user to trigger the reconstruction process manually.

\subsubsection{Insert Operation}
After a certain point, the accuracy of a learned index will inevitably deteriorate due to the continuous insertion of new objects that exhibit distribution shifts, or deletion of existing objects. Furthermore, adapting a trained model within a learned index to a new distribution results in a phenomenon called catastrophic forgetting~\cite{mccloskey1989catastrophic,kirkpatrick2017overcoming} in which the model forgets the already learned knowledge. A subfield of machine learning called continual learning~\cite{wang2024comprehensive} studies methods that allow models to adapt to situations where the learned knowledge is to be adjusted throughout the model's existence. Such methods attempt to balance the trade-off between learning plasticity, i.e., how well the new task is learned, and memory stability, i.e., how well the model remembers previous knowledge. 

\subsubsection{Selecting the Model Architecture}
Choosing the optimal model architecture requires machine learning expertise and can be computationally expensive as well as time-consuming. In this paper, we do not focus on the internal model tuning. However, we presume that an automated search for a model architecture~\cite{ren2021comprehensive,elsken2019neural} would lead to further performance improvements.

\subsection{Dynamicity in Traditional Indexes}

The feasibility of dynamizing an index structure depends on the index type -- while some naturally accommodate dynamic operations, others present inherent challenges.

\subsubsection{Tree-based}
Tree-based methods hierarchically divide the search space so that each node covers a given sub-space, thereby increasing the specificity
of the search with deeper tree levels. While some tree indexes, particularly those based on a pivoting mechanism~\cite{yianilos1993data, DGSZ03} 
are inherently dynamic, others~\cite{muja2014scalable, bernhardsson2015annoy} either presume a static use case or require a balanced structure for optimal performance, requiring a re-balancing operation to prevent the influx of new data from degrading the performance.

\subsubsection{Graph-based}
Graph-based indexes~\cite{wang2021comprehensive} are considered the state of the art in approximate nearest neighbor search~\cite{aumuller2020ann}. Conceptualized as proximity graphs~\cite{fu2017fast,malkov2014approximate}, they approximate their theoretical counterparts~\cite{fortune2017voronoi,toussaint1980relative}, reducing building costs and accelerating search by omitting non-relevant edges. HNSW~\cite{malkov2018efficient}, the most researched index of this category, is incorporated into a wide range of vector databases~\cite{pan2024survey} and is dynamic by design due to its incremental building procedure, although without a delete operation\footnote{\url{https://github.com/nmslib/hnswlib/issues/4}}.


\subsubsection{LSH-based}
Most LSH-based methods~\cite{indyk1998approximate,gionis1999similarity,lv2017intelligent,liu2014sk} are built for static use and thus fail to efficiently accommodate dynamic data loads. Adding new data to a created LSH typically requires recalculating the hash functions and updating the index. Two notable works attempt to extend the static nature of LSH to dynamic use cases: LSH forest~\cite{bawa2005lsh} through the use of a multiple prefix tree structure and PLSH~\cite{sundaram2013streaming} by employing an insert-optimized hash table.


\subsection{Dynamization Frameworks}
The central theme of this paper is the transformation of a static index into a dynamic one -- dynamization -- through the use of simple, index-agnostic operators. The goal of dynamization, as proposed in the original Bentley-Saxe method~\cite{bentley1980decomposable}, is to abstract away from the details of particular indexing solutions by integrating procedures for insertion and deletion as well as transforming a single index into a series of indexes of progressively larger sizes. Although this method has been applied in numerous use cases~\cite{naidan2012static, rumbaugh2023practical, xie2021spatial}, it struggles with deletion performance, limited support for various query types and a lack of generality beyond decomposable search problems. Recent work~\cite{rumbaugh2024towards} addressed the shortcomings of Bentley-Saxe by proposing a more general dynamization framework, paving the way for modern, robust index dynamization.

As opposed to~\cite{rumbaugh2024towards}, the dynamization method presented in this paper is not an extensive ready-to-use framework. Instead, we introduce a minimalistic approach consisting of two extension operations and a basic set of restructuring policies that can be implemented into any partitioning-based index. While comprehensive frameworks are suitable for complex dynamization needs (e.g., performance tuning), in this paper, we show that many use cases can be addressed with our lightweight approach.


\section{Methods}
\label{sec:methods}
The static index we have chosen for dynamization is a hierarchical learned index called the Learned Metric Index (LMI)~\cite{antol2021learned}. The model was first conceptualized as a tree structure composed of learned models, where the root node is a model trained on all the data, with a pre-defined number of classification categories corresponding to its children. The child nodes are either leaf nodes, i.e., buckets containing the given subset of the data, or inner nodes, which correspond to another learned model -- this model once again partitions its given subset of data into a pre-defined number of classes.\footnote{Specifically, in the index we dynamized, the single predictive unit is a Multi-Layer Perceptron (MLP) with one hidden layer of 128 neurons. Prior to training the model, each data point was assigned a category by the K-Means clustering algorithm -- the MLP was then trained with a classification objective in a supervised way.} Once such an index is built, it can be used to resolve search queries by recursively classifying the query objects until a certain number of leaf nodes has been reached.

\subsection{Dynamized Learned Index}
\label{sec:dynamized}

For a tree-like structure to become dynamic, we need to define mechanisms for adaptive expansion of its nodes. For a node $n$ which has reached its defined capacity (thereby requiring an update), we describe two basic extension mechanisms: node split (deepening) and node expansion (broadening).

\begin{figure}[t!]
  \centering
  \includegraphics[width=0.9\columnwidth]{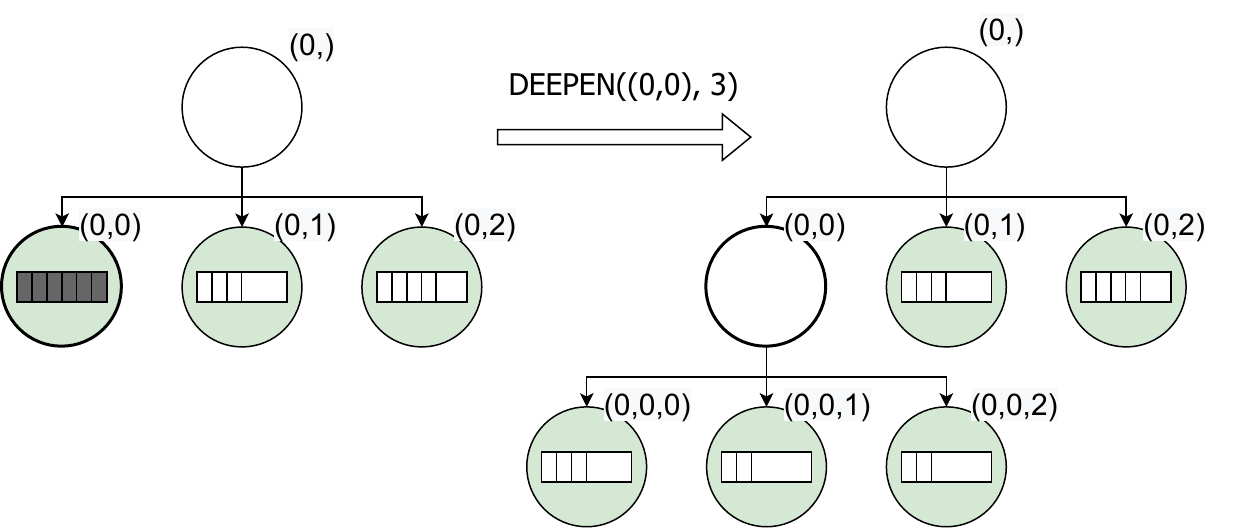}
  \caption{Overview of the deepening operation.}
  \label{fig:deepen}
\end{figure}

\textit{Deepening} is defined on a leaf node -- upon reaching maximum capacity, the node is transformed into an inner node, and its objects are dispersed into newly created child nodes, as illustrated in Figure~\ref{fig:deepen}. In the context of a learned index, deepening would be equivalent to creating a new model and training on the node's objects with a given target number of child nodes -- see Algorithm~\ref{alg:deepen}.

\begin{algorithm}[ht]
\SetAlgoLined
\textbf{Input:} structure $S$, a data node $n$, \# of new children $n\_child$, distance function $d$ \\
\textbf{Output:} modified structure $S$ \\
labels = cluster(n.objects, n\_child, d) \\
model, positions = Model(n.objects, labels, n\_child) \\
new\_n = InnerNode(n.pos, model) \\
S.insert\_node(new\_n) \\
S.insert\_child\_nodes(n\_child, new\_n.pos, n.objects, positions) \\
delete\_nodes([n]) \\
S.check\_consistency() \\
\textbf{return} S;
\caption{Deepen}
\label{alg:deepen}
\end{algorithm}

\vspace{-0.3cm}

\begin{figure}[t!]
  \centering
  \includegraphics[width=\columnwidth]{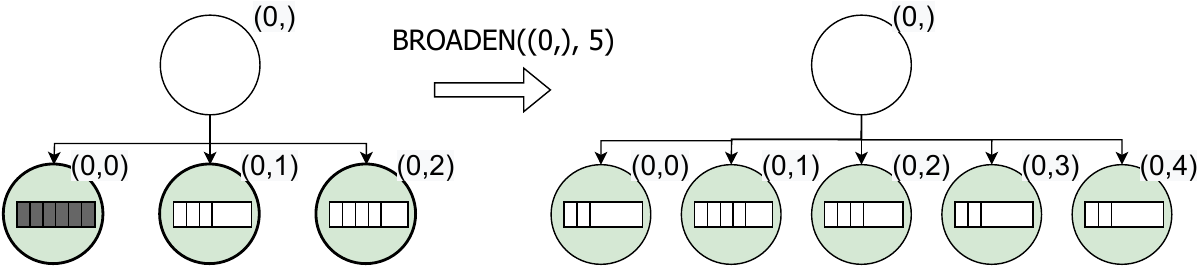}
  \caption{Overview of the broadening operation.}
  \label{fig:retrain}
\end{figure}

While deepening triggers vertical growth, \textit{broadening} extends the index horizontally to avoid an unnecessarily deep structure. Broadening is defined as the re-creation of a node (either inner or leaf) from scratch with its current objects. In learned indexing, this involves re-partitioning and retraining of the learned model with all the relevant objects (potentially also objects on the grandchildren's level). This operation is fully described in Algorithm~\ref{alg:retrain} and visualized in Figure~\ref{fig:retrain}. Note that while the complete recreation of a node may not strictly meet the definition of "broadening", as the original node technically ceased to exist, in the context of learned indexing, adding new categories to a model after it has been built would lead to performance deterioration due to catastrophic forgetting (as mentioned in Section~\ref{sec:relatedwork}).

\begin{algorithm}[ht]
\SetAlgoLined
\textbf{Input:} structure $S$, an inner node $n$, \# of new children $n\_child$, distance function $d$ \\
\textbf{Output:} modified structure $S$ \\
data\_n, inner\_n = S.find\_nodes\_in\_subtree(n.pos) \\
objects = collect\_objects(data\_n) \\
labels = cluster(objects, n\_child, d) \\
model, positions = Model(n.objects, labels, n\_child) \\
n.model = model \\
new\_n = InnerNode(n.pos, model) \\
S.insert\_child\_nodes(n\_child, new\_n.pos, objects, positions) \\
delete\_nodes(data\_n + [inner\_n]) \\
S.check\_consistency() \\
\textbf{return} S;
\caption{Broaden}
\label{alg:retrain}
\end{algorithm}

\begin{figure}[t!]
  \centering
  \includegraphics[width=0.8\columnwidth]{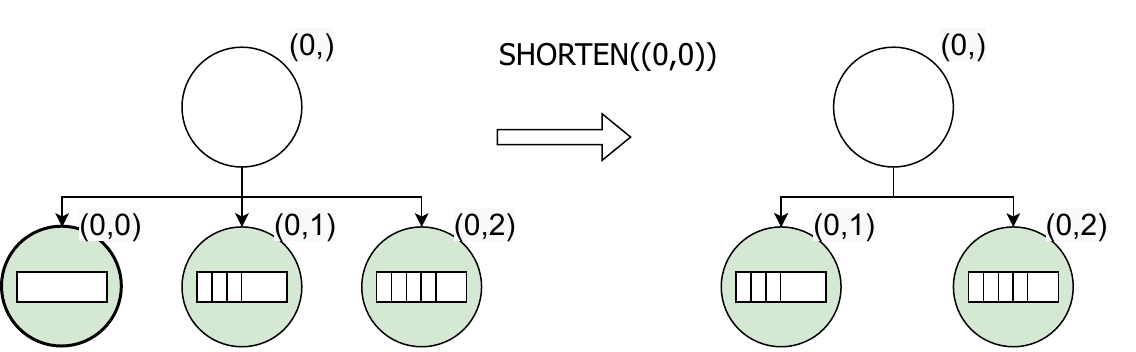}
  \caption{Overview of the shortening operation.}
  \label{fig:shorten}
\end{figure}

\textbf{Removal of a defective or obsolete node.} 
In case of a severe underflow of a data node or other type of defect, this node should be deleted and its objects (if any) reinserted using the \textit{shorten} operation. This operation is implemented by removing a category from among the leaf nodes, which in our case involves removing one of the learned model's output neurons. Unlike the insertion of an additional neuron, this removal does not affect the model's prediction mechanism. For more details, see Algorithm~\ref{alg:shorten} and Figure~\ref{fig:shorten}. 

\textbf{Policies that trigger the structural changes.} After defining the basic operations for a structural update, the next step is to establish policies to invoke them. First, it is reasonable to define minimum and maximum bounds on the leaf node capacity to impose limits on the costs of sequential search. Similarly, such bounds should be defined on the inner nodes for optimal discriminative power of a single model. We chose to define a policy of detecting and resolving any violations of these bounds -- in case of a leaf node having less than 5 objects (\textit{underflow}), we immediately delete it and redistribute its objects among the other nodes. To manage \textit{overflow}, the structure ensures that the average occupancy of all leaf nodes remains below 1\,000. When this bound is violated, nodes with the highest occupancy are extended until the average falls back below the threshold. In terms of depth, shallow indexes are preferred -- for the 1\,000\,000 objects in our experimental dataset, it only grows up to a maximum of two levels.

\begin{algorithm}[ht]
\SetAlgoLined
\textbf{Input:} structure $S$, an array of data nodes $arr_n$ \\
\textbf{Output:} modified structure $S$ \\
Initialize empty array A \\
\For{each item $n$ in $arr_n$} {
    parent = n.get\_parent() \\
    parent.remove\_child(n.pos) \\
    A.add(n.objects)
}
S.insert($A$);
S.delete\_nodes($arr_n$);
\textbf{return} S;
\caption{Shorten}
\label{alg:shorten}
\end{algorithm}

\subsection{Evaluation Baseline}
A dynamized index is not necessarily an improvement in terms of total costs -- while it should theoretically scale better with growing datasets, it introduces additional overhead not present in the static version. 
To determine the conditions under which dynamization is worthwhile, the dynamized index should be evaluated against its static counterpart.

One way to do this is to simply keep inserting objects into the static index and compare it with the growing adaptive version. This will necessarily lead to deterioration of the quality of the query results in the static index, but the static version will only require a single build, likely leading to lower overall build costs. In the experimental evaluation, we will refer to this as the \textbf{No rebuild} baseline.

Another way is to take advantage of the fact that any static index can be made somewhat dynamic 
by rebuilding it in its entirety after a certain threshold of new objects, thus allowing it to adapt to new data at the cost of additional build costs -- we will refer to this as the \textbf{Naive rebuild} baseline. Implementing the Naive rebuild method involves the selection of a single parameter which determines how often the index is discarded and rebuilt from scratch. 
Let us call this parameter the \textit{rebuild interval}, and define it in terms of the number of new objects that can be added before the index is rebuilt (e.g., a rebuild interval of 10\,000 means that after we build the index, the next 9\,999 objects are simply added to the structure as is, and the 10\,000th object triggers a full rebuild). In the next section, we will discuss the selection and optimization of this parameter.

\subsection{Amortized Cost Model}
The typical problem when evaluating static and dynamic indexes is the fact that their costs are distributed very differently -- some indexes trade good query performance for high upfront costs of building a well-performing index, while others may be able to start answering queries quickly by starting with an imperfect structure that adapts over time. Thus, if build time is included in the evaluation, the static index is greatly disadvantaged at the beginning, as it takes a long time to even answer the first query. If the build time is excluded, however, various adaptive methods may be punished for not taking the time to build the perfect structure ahead of time.

A typical approach in research evaluation is to present two sets of plots, comparing the build costs and search costs side by side (for example in~\cite{lampropoulos2023adaptive}). While this can help identify outliers or unusual behavior (such as an index with exorbitant build costs for the sake of minor query improvement), it does not provide a clear overall picture of the actual trade-off between build and search costs. The information is dispersed across two separate sets of plots, often on different scales, making it hard to assess directly.

An ideal evaluation metric should combine the search costs and build costs into a single objective that can be compared on a per-query basis regardless of how dynamic the method is. Thus, in addition to the search costs of a query, which are easily determined, we should also consider the amortized build costs of that query -- that is, the total build costs of a structure divided by the number of queries performed on that structure.

However, the number of queries can only be determined once we define a new property of the system: the relative number of queries per inserted object. We will call this the \textit{querying frequency}, and define it simply as $\frac{\# queries}{\# new\_objects}$. We will assume querying frequency to be an inherent property of the given indexing scenario, which does not change during the lifetime of the database. This allows us to relate the number of new objects added to the database (which dictate the rebuild costs of a dynamic structure) with the number of queries performed on the database (which determine the amortization).

The only other factor that determines whether structural adjustments are worth their costs is how accurate the queries need to be -- the \textit{target recall}. If lower query recall is sufficient, deterioration of the structure is less problematic, and rebuilds are not as worthwhile. If high recall is needed, the index needs to be maintained in peak condition via more frequent rebuilds -- deterioration of query quality is more punishing, so even more costly rebuilds are justified.

Once we determine the querying frequency of our indexing scenario and the desired (average) recall of our queries, we can infer the amortized search cost as:

\large
\begin{equation*}
    \texttt{AC} = \texttt{SC} + \frac{\texttt{BC}}{\texttt{RI}*\texttt{QF}}
\end{equation*}
\normalsize

\noindent where \texttt{AC} is the amortized cost, \texttt{SC} is the search cost of a single query (i.e., how many seconds it takes for an average query to achieve the target recall), \texttt{RI} is the rebuild interval of the given index (in terms of the average number of new objects that trigger a rebuild), \texttt{BC} is its build cost (in seconds), and \texttt{QF} is the querying frequency of the given indexing scenario (in terms of queries per insert).

In other words, if an index is rebuilt after adding 1\,000 new objects, and the querying frequency is 100 queries per new object, then one build of the index will last for 100\,000 queries. Therefore, the amortized cost of a query should include $\frac{1}{100000}$ of the index build costs in addition to the immediate search costs.

This allows methods with infrequent but costly rebuilds to be directly compared with methods that perform gradual, less extensive updates.

\subsubsection{Optimal Rebuild Interval}

The concept of amortized cost can also be used to optimize the \textit{Naive rebuild} baseline method described earlier -- if we can determine the indexing scenario ahead of time (i.e., estimate the typical number of new queries per insert and decide on the desired recall of the database), we can optimize when a full rebuild of the structure should be triggered.

\begin{figure}
    \centering
    \includegraphics[width=0.93\linewidth]{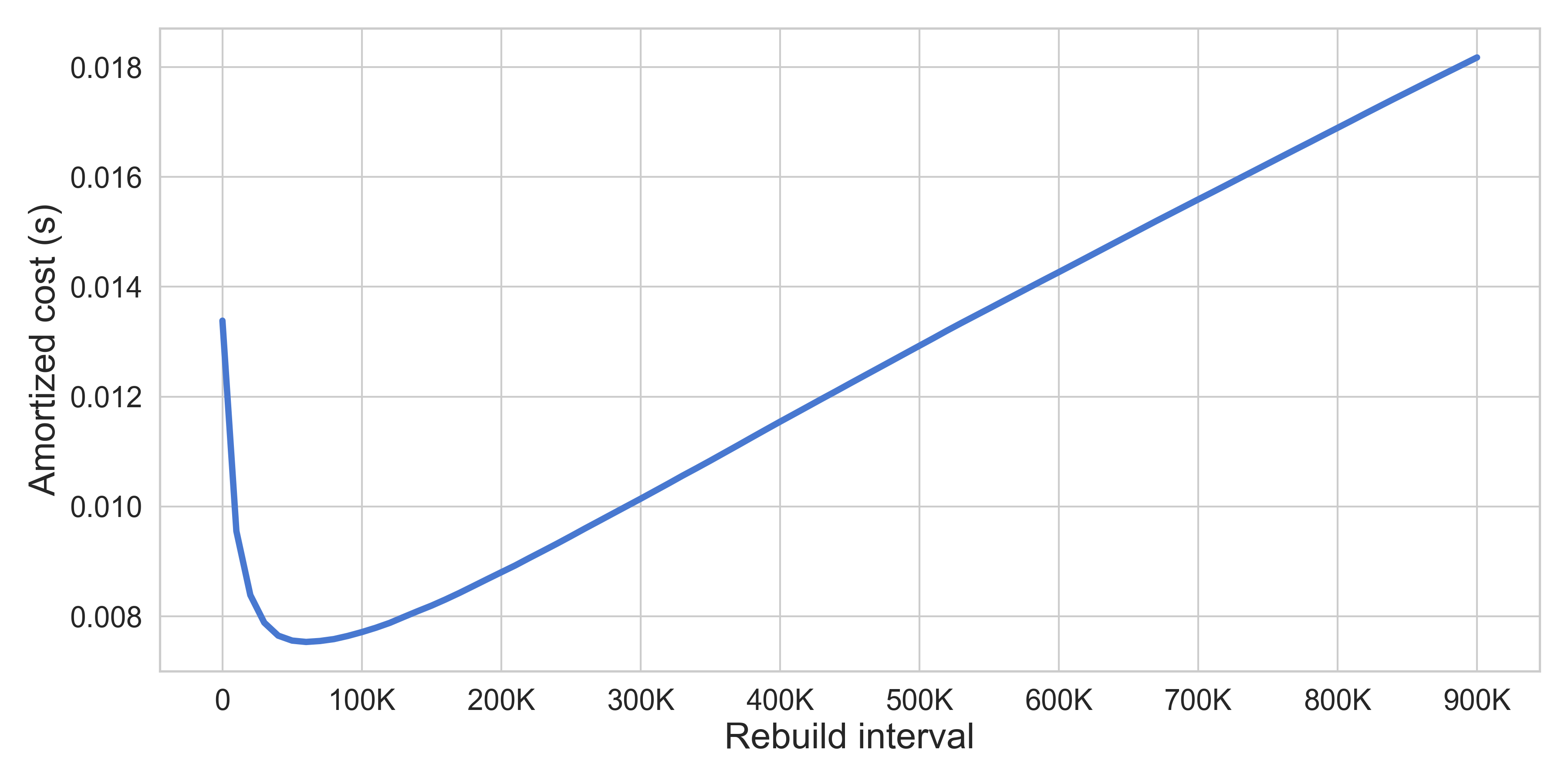}
    \caption{The amortized cost of a Naive rebuild baseline at different rebuild intervals (setup with 1 query per new object and a target recall of 0.5).}
    \label{fig:AC_plot}
\end{figure}

The optimal rebuild interval must balance the cost of rebuilding the entire structure against the increasing search costs of the average query caused by the gradual deterioration of the index. As objects are added without rebuilding the index, the per-query build costs decline over time, since there are more queries to divide the original build cost. On the other hand, the per-query search costs increase over time as the index deteriorates due to a lack of a structural update. Thus, the sum of these two components inevitably creates a single optimum, representing the ideal rebuild interval with the perfect balance of search and build costs that minimizes the amortized cost per query. Figure~\ref{fig:AC_plot} illustrates how the amortized cost per query relates to the rebuild interval in a particular indexing scenario, derived from one of the experimental cases discussed in the next section. 

For additional details on how the optimal rebuild interval varies based on the database size and indexing scenario, and the relative influence of search versus build costs, please refer to Tables~\ref{tab:rebuild_intervals} and~\ref{tab:search_cost_percentage} in the Appendix.

By definition, the rebuild interval can only be optimized for a specific scenario, that is, a single combination of query frequency and target recall. If these properties cannot be accurately determined in advance, or if they change over time, the performance of this approach may end up being far from optimal. The penalty for a suboptimal selection of the rebuild interval parameter will be evaluated in the next section.

\section{Experiments}

For our experimental evaluation, we use SIFT descriptors~\cite{lowe2004distinctive}, one of the datasets used in ANN-benchmarks~\cite{aumuller2020ann}. This dataset consists of 1 million objects of 128 dimensions with the Euclidean distance metric, and additionally contains 10\,000 queries in a 30-NN setup for experimental evaluation.

Since the amortized cost metric calculation depends on two variables -- the number of \textit{queries per inserted object} (\texttt{QPI}) and \textit{target recall} (\texttt{TR}) -- the evaluation considers scenarios with two extreme settings of either variable, for a total of 4 scenarios. We set a high querying frequency at 100 queries per insert (corresponding, for instance, to a typical social media feed), and low querying frequency at 1 query per insert (corresponding to a monitoring service or a messaging system). For target recall, we set the extremes at 0.9 (high) and 0.5 (low) -- in the context of a 30-NN query, this means that we require the database to return, on average, at least 27 objects at high target recall and 15 objects at low target recall.


As described in Section~\ref{sec:methods}, we compare the dynamized version of the learned index against two baselines -- the \textit{Naive rebuild} baseline, which rebuilds the entire index periodically based on a pre-defined rebuild interval, and the \textit{No rebuild} baseline, which simply stores incoming objects without adjusting its structure until it runs out of experimental data to process. We performed the experiment with various database sizes, ranging from 100\,000 to 900\,000. For the baselines, this means that the database was first built using the given number of initial objects, and then new objects were added -- in the case of the \textit{Naive rebuild} baseline, the objects were added until a rebuild was triggered, in the \textit{No rebuild} baseline, they were added until all 1\,000\,000 experimental data objects were indexed. The dynamized index was built gradually starting with no objects, and its amortized performance was simply evaluated after every 100\,000 objects as the index grew and adapted.

Since the \textit{Naive rebuild} baseline is sensitive to the selection of rebuild interval, we optimized this parameter for each of the 4 experimental scenarios, resulting in 4 different performance curves of this method. As a result, in each evaluated scenario, one of the Naive rebuild baselines shows the optimal solution (since the rebuild frequency is set up perfectly for that particular scenario), and the other Naive rebuild baselines show the penalty for parameterizing the method incorrectly.

As for the \textit{No rebuild} baseline, since the index is only built once (using the initial number of objects), the build costs are as low as possible for the given initial database size, but the quality of queries will keep deteriorating towards exhaustive search in the limit. Due to the size of our experimental dataset (1\,000\,000 objects), we can only simulate smaller and smaller increments of this deterioration as we use more objects to build the initial database. In other words, when the initial database size is 100\,000, the plotted results show how much queries deteriorate if database size increases by 900\% with no rebuild; when we start with 900\,000 objects, the plot shows how much performance deteriorates if we add $\sim$11\% additional objects, as there are no more objects to insert. Thus, when plotting the amortized cost, the results of this baseline keep converging towards the dynamized methods, not due to improved performance of the method, but because the fixed size of the experimental dataset limits the ability to model larger-scale growth. 


Note that the static index used in the \textit{Naive rebuild} and \textit{No rebuild} baselines is a single-level structure, implemented as a single MLP, and is parameterized to hold an average of 1\,000 objects per bucket. Additionally, note that only insertion of new objects is considered -- deletion and re-balancing can also be implemented in a straightforward manner (as shown in Section~\ref{sec:dynamized}), but the experimental evaluation of such operations is beyond the scope of this paper. 


\begin{figure}[h!]
    \centering
    \includegraphics[width=0.93\linewidth]{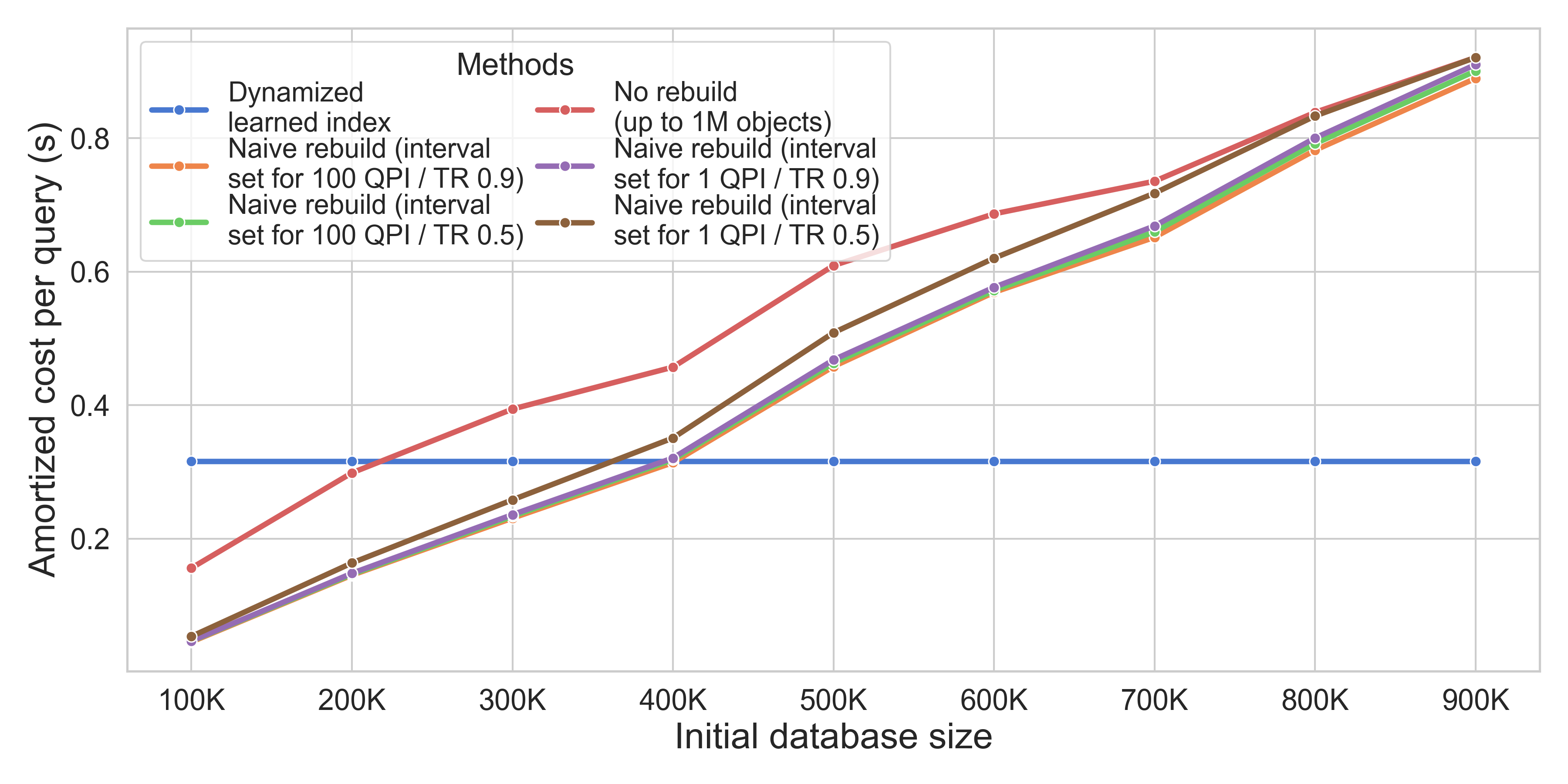}
    \vspace{-0.5cm}
    \caption{Amortized costs of the dynamized index and various baselines in the high intensity---high target recall scenario (100 queries per insert, target recall of 0.9).}
    \label{fig:highQI_highTR}
\end{figure}

\begin{figure}[p]
    \centering
    \includegraphics[width=0.93\linewidth]{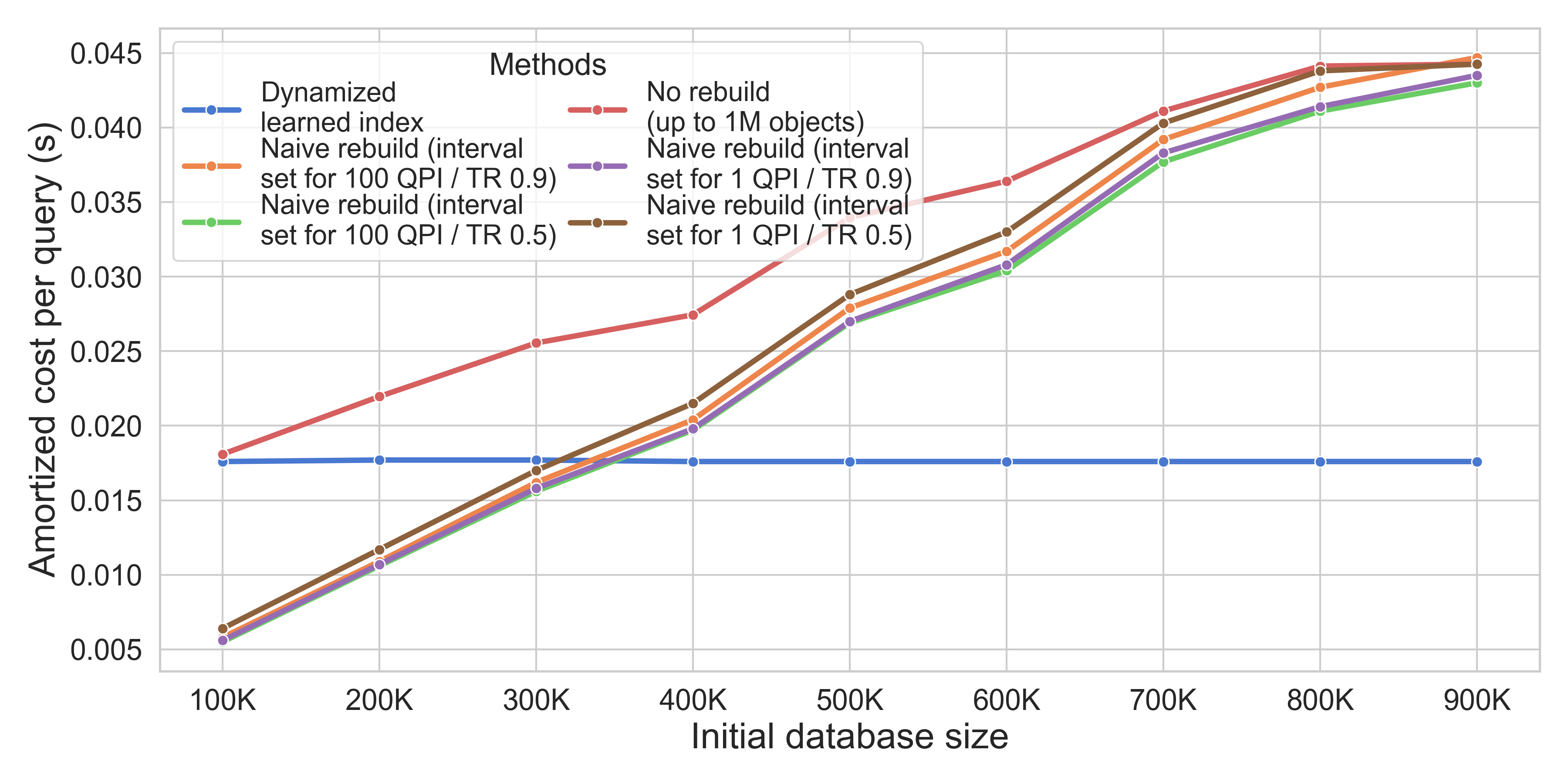}
    \vspace{-0.7cm}
    \caption{Amortized costs of the dynamized index and various baselines in the high intensity---low target recall scenario (100 queries per insert, target recall of 0.5).}
    \label{fig:highQI_lowTR}
\end{figure}

\begin{figure}[p]
    \centering
    \includegraphics[width=0.93\linewidth]{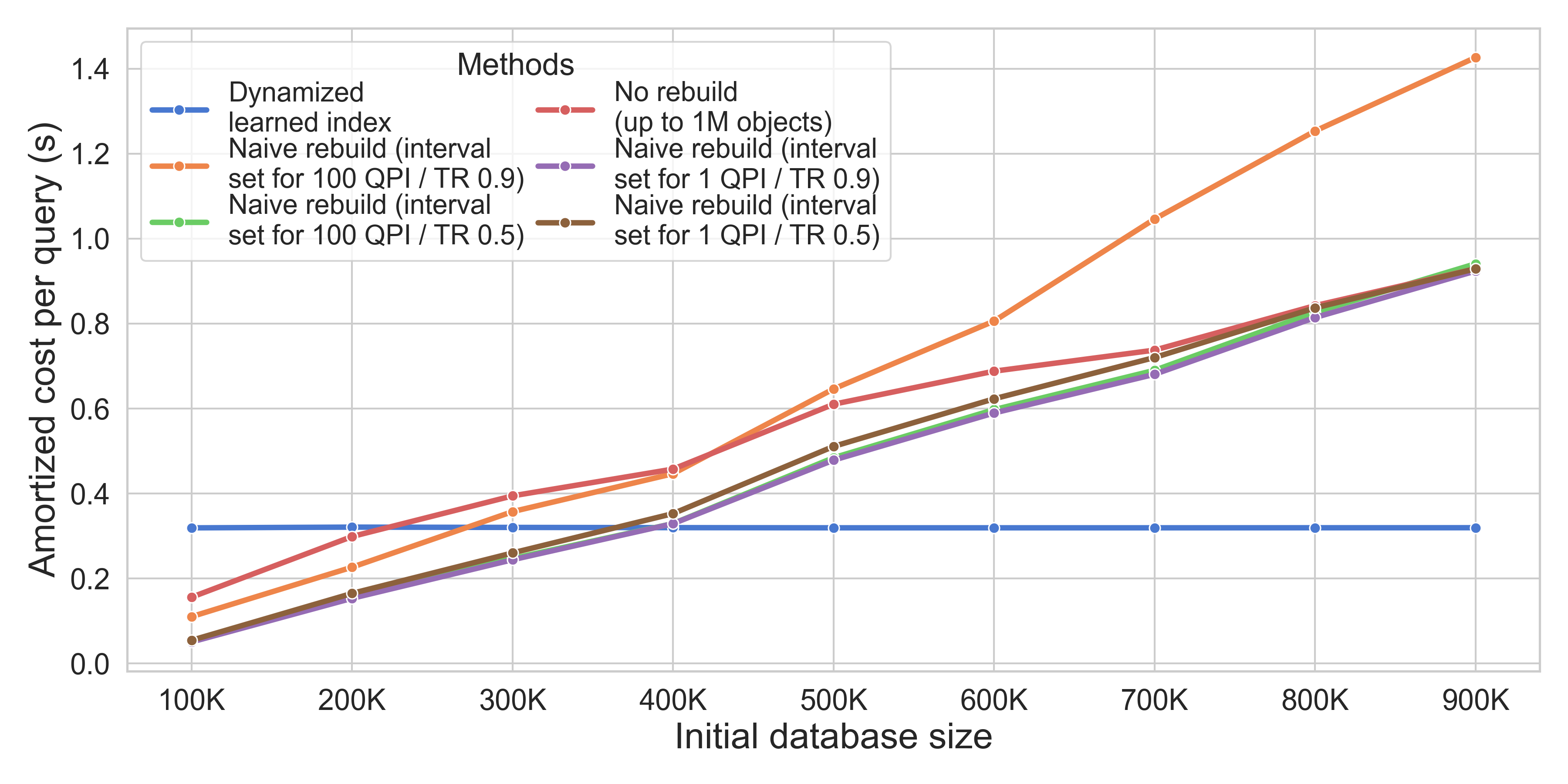}
    \vspace{-0.7cm}
    \caption{Amortized costs of the dynamized index and various baselines in the low intensity---high target recall scenario (1 query per insert, target recall of 0.9).}
    \label{fig:lowQI_highTR}
\end{figure}

\begin{figure}[p]
    \centering
    \includegraphics[width=0.93\linewidth]{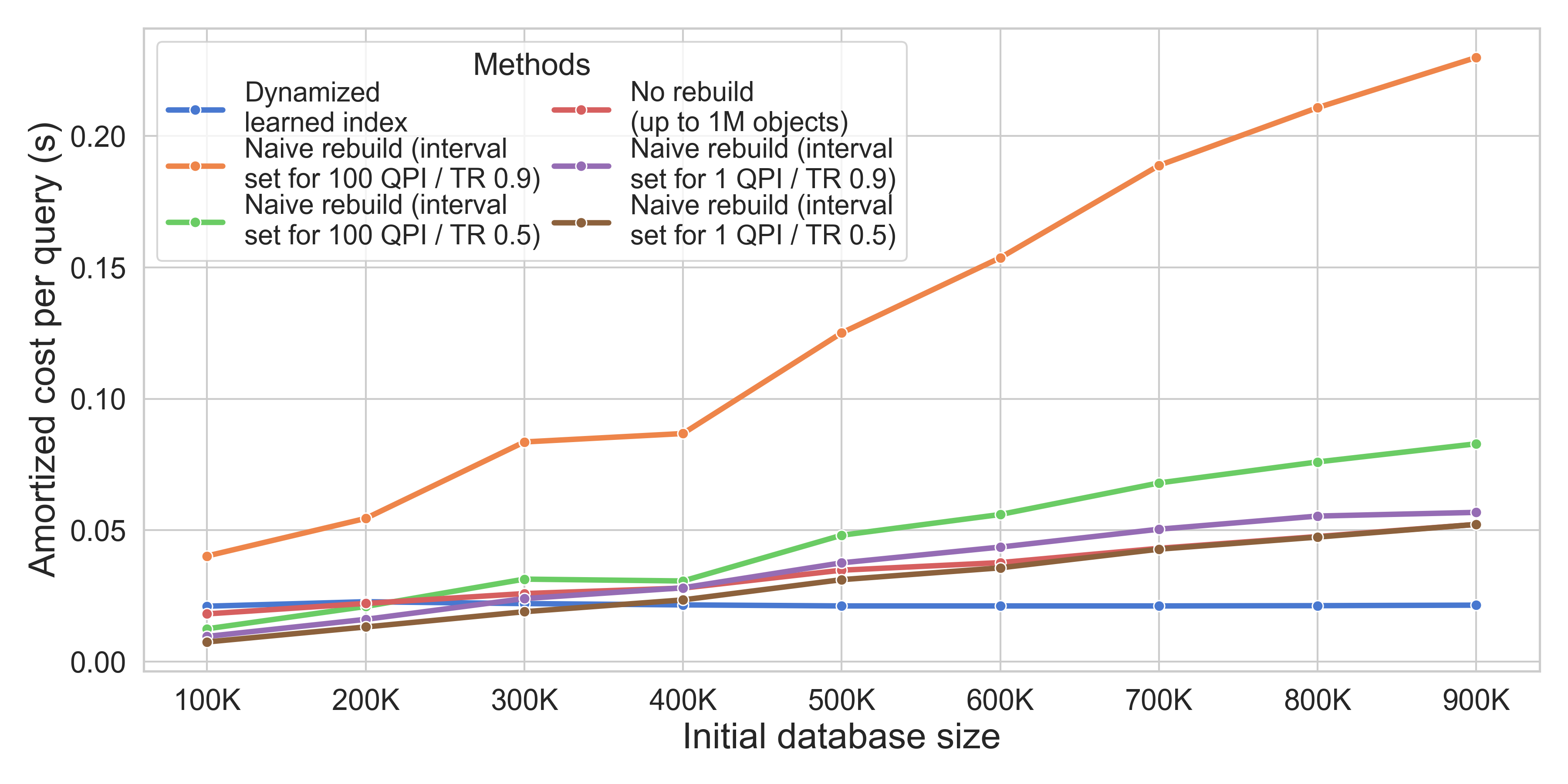}
    \vspace{-0.7cm}
    \caption{Amortized costs of the dynamized index and various baselines in the low intensity---low target recall scenario (1 query per insert, target recall of 0.5).}
    \label{fig:lowQI_lowTR}
\end{figure}

\clearpage

Figures~\ref{fig:highQI_highTR}-\ref{fig:lowQI_lowTR} show how amortized costs scale with the database size for each combination of high/low querying frequency and high/low target recall. Keep in mind that the "Initial database size" label only holds true for the baseline methods -- the dynamized index always has an initial database size of 0, and it is merely evaluated at various size thresholds as it grows.

Our first observation is that the plots communicate the overall costs of a given indexing approach in a very transparent manner, making it clear which methods perform best under which conditions, and how this is affected by the growing datasets.
As expected, the amortized costs of the \textit{Naive rebuild} and \textit{No rebuild} baselines keep increasing with database size -- as a result, the dynamized method inevitably performs better on larger datasets. 

In scenarios with higher querying intensity (100 queries per object), the various rebuild intervals of the \textit{Naive rebuild} method do not seem to make much of a difference. In lower intensity scenarios, however, the method with the most frequent rebuilds (parameterized for high intensity scenarios) is severely punished, usually performing worse than a method that performs no rebuilds at all. The reason for this can be inferred from Figure~\ref{fig:AC_plot} -- if the rebuild interval is too small, the penalty in terms of amortized costs is much steeper than if the interval is too large. Aside from this extreme case, the parameterization of the \textit{Naive rebuild} method is not as impactful as we had expected -- the choice between the naive and dynamized approaches is usually much more important than the choice of the rebuild interval between different \textit{Naive rebuild} variants.

\section{Acknowledgments}

This work was supported by the Ministry of Education, Youth and Sports of the Czech Republic (ELIXIR-CZ, Grant No. LM2023055), e-INFRA CZ (ID:90254), and the Czech Science Foundation [GF23-07040K].

\section{In Conclusion}
We introduced a method for dynamizing a static index, using a streamlined set of policies and operations to enable adaptive expansion. This approach was used to transform a previously static learned index for complex datasets into a dynamic one. In addition, we introduced an amortized cost model to evaluate the benefits of the dynamic version against its static counterpart in various scenarios. As expected, the results show that the dynamized index scales very well with database size in all scenarios, and while its overhead may not be justified for smaller databases, the dynamization quickly proves advantageous as the database grows. 

Taken together, our dynamization approach offers a straightforward way to transition from static to dynamic learned indexes, significantly broadening their applicability in complex data, while the evaluation method provides a practical way to assess the trade-offs of dynamization and identify scenarios where its benefits outweigh its costs.

\clearpage

\bibliographystyle{splncs04}
\bibliography{citations}

\newpage
\section*{Appendix}

The following tables detail the Naive rebuild method parameterized for various indexing scenarios. Table~\ref{tab:rebuild_intervals} shows the optimal number of new objects that should trigger a full rebuild of the index. As a reminder, \texttt{QPI} refers to the number of \textit{queries per insert} and \texttt{TR} refers to \textit{target recall} in a given scenario.

The rebuilds are most frequent in a scenario with high querying intensity and high target recall, since this scenario is most sensitive to deterioration in search quality -- it is hard to reach high recall in a decayed index, and since there are many queries, the penalty for slow searching quickly outpaces the costs of rebuilding the index from scratch. The opposite is true for the low querying intensity---low target recall scenario, where the penalty for deteriorating search quality is relatively mild -- it is still feasible to reach a recall of 0.5 in a low-quality structure, and since there are few queries, it takes a long time for the slower search to incur so much overhead that a full rebuild becomes worthwhile.

The single N/A value in each table is due to the limited number of objects in the dataset -- if the initial database size is 900\,000 but the rebuild interval would be higher than 100\,000, we run out of objects before even a single index rebuild could be triggered and evaluated.

\vspace{-0.2cm}

\begin{table}[]
    \centering
    \begin{tabular}{c||c|c|c|c}
         \multirow{2}{*}{DB size} & 100 QPI & 100 QPI & 1 QPI & 1 QPI \\
         & TR 0.9 & TR 0.5 & TR 0.9 & TR 0.5 \\
         \hline
         100\,000 & \ 2\,700 & \ 7\,900 & 29\,900 & \ 70\,000 \\
         200\,000 & \ 4\,300 & 11\,500 & 30\,000 & 110\,000 \\
         300\,000 & \ 4\,800 & 16\,500 & 30\,000 & 120\,000 \\
         400\,000 & \ 6\,000 & 20\,600 & 40\,000 & 160\,000 \\
         500\,000 & \ 5\,900 & 23\,600 & 40\,000 & 180\,000 \\
         600\,000 & \ 6\,000 & 29\,700 & 40\,000 & 190\,000 \\
         700\,000 & \ 8\,000 & 30\,000 & 50\,000 & 240\,000 \\
         800\,000 & 10\,600 & 30\,000 & 50\,000 & 190\,000 \\
         900\,000 & \ 6\,600 & 30\,000 & 60\,000 & N/A
    \end{tabular}
    \vspace{0.15cm}
    \caption{The optimal rebuild intervals for the Naive rebuild method in different indexing scenarios.}
    \label{tab:rebuild_intervals}
\end{table}

\vspace{-1.5cm}

\begin{table}[]
    \centering
    \begin{tabular}{c||c|c|c|c}
    \multirow{2}{*}{DB size} & 100 QPI & 100 QPI & 1 QPI & 1 QPI \\
         & TR 0.9 & TR 0.5 & TR 0.9 & TR 0.5 \\
         \hline
         100\,000 & 98.53\% & 97.23\% & 89.29\% & 84.77\%  \\
         200\,000 & 99.08\% & 97.30\% & 96.46\% & 88.75\%  \\
         300\,000 & 99.19\% & 97.75\% & 96.59\% & 89.06\%  \\
         400\,000 & 99.34\% & 97.97\% & 97.45\% & 91.10\%  \\
         500\,000 & 99.32\% & 98.01\% & 97.76\% & 92.37\%  \\
         600\,000 & 99.36\% & 98.33\% & 97.80\% & 92.35\%  \\
         700\,000 & 99.51\% & 99.19\% & 98.20\% & 94.04\%  \\
         800\,000 & 99.63\% & 99.14\% & 98.27\% & 92.18\%  \\
         900\,000 & 99.38\% & 99.06\% & 98.55\% & N/A  \\
    \end{tabular}
    \vspace{0.15cm}
    \caption{The percentage of search costs within the overall amortized costs per query.}
    \label{tab:search_cost_percentage}
\end{table}

\end{document}